\documentclass[10pt,letterpaper,twocolumn]{article} 
\usepackage{OL}
\usepackage[draft]{hyperref} 
\usepackage{graphicx}

\def\kv{{\bf k}}
\def\Kv{{\bf K}}

\def\qv{{\bf q}}
\def\rv{{\bf r}}
\def\Dk{{\bf \Delta k}}
\def\dk{{\Delta k}}

\begin{document}

\twocolumn[ 

\title{Experimental Confirmation of the General Solution to the
  Multiple Phase Matching Problem}

\author{Alon Bahabad,$^\dag$ Noa Voloch,$^\ddag$ Ady Arie,$^\dag$ and
  Ron Lifshitz$^\ddag$}

\address{$^\dag$School of Electrical Engineering, Wolfson Faculty of
  Engineering, Tel Aviv University, Tel Aviv 69978, Israel\\
  $^\ddag$School of Physics and Astronomy, Raymond and Beverly Sackler
  Faculty of Exact Sciences, Tel Aviv University, Tel Aviv 69978,
  Israel}


\begin{abstract}
  We recently described a general solution to the phase matching
  problem that arises when one wishes to perform an arbitrary number
  of nonlinear optical processes in a single medium \protect{[}{\it
    Phys.\ Rev.\ Lett.} {\bf 95}, 133901 (2005)\protect{]}. Here we
  outline in detail the implementation of the solution for a one
  dimensional photonic quasicrystal which acts as a simultaneous
  frequency doubler for {\it three\/} independent optical beams. We
  confirm this solution experimentally using an electric field poled
  KTiOPO$_4$ crystal. In optimizing the device, we find---contrary to
  common practice---that simple duty cycles of 100\% and 0\% may yield
  the highest efficiencies, and show that our device is more efficient
  than a comparable device based on periodic quasi-phase-matching.
\end{abstract}


] 

\section{Introduction}

Three-wave mixing is a nonlinear optical process that can take place
within dielectric materials having a nonlinear $\chi^{(2)}$ coupling
coefficient. Such processes are used for a variety of optical
frequency conversion applications. Usually, due to dispersion, the
three interacting beams do not propagate in phase and so efficient
energy transfer between them is prevented.\cite{Armstrong1962} One of
the common methods to solve this problem, called quasi-phase-matching
(QPM), is to periodically modulate the sign of the relevant component
of the nonlinear dielectric tensor at a period corresponding to the
phase mismatch\cite{Armstrong1962,Fejer_QPM1992}. This approach is
very successful, but unless one is extremely lucky it is limited to
the phase matching of a single optical process. In recent years, the
need to simultaneously phase match several different processes arose
in numerous applications such as creation of multiple radiation
sources\cite{Ming_THG1997}, of multi-colored
solitons\cite{Kivshar_CascadingSoliton1999}, of multi-partite
entanglement sources,\cite{Pfister_TripleCoincidence_OptLett2005} and
for all-optical processing.\cite{Stegman_Cascading1996} This need was
addressed by developing {\it ad hoc\/} generalizations for the
quasi-phase-matching procedure, based either on
periodic  structures in one dimension\cite{Ellenbogen} (for non-collinear processes) and two dimensions\cite{Berger_NPC1998,%
  Broderick_Hexagonal2000,OQE2007} or specific quasiperiodic
structures in one\cite{Keren_PRL2001,Ming_THG1997,Ming_MultipleSHG2002,%
  Ming_FourColor2004,Fejer_MultipleHarmonics1999} and
two\cite{Bratfalean_NPQC2005} dimensions. In a recent
publication\cite[henceforth LAB]{Lifshitz_PRL2005} we explained how to
solve the most general problem of multiple phase matching---designing
a device to phase match an arbitrary set of processes, defined by any
given set of phase-mismatch values. The LAB solution is based on the
general observation that the phase matching problem is a consequence
of momentum conservation, and that in crystalline matter, {\it i.e.}
matter with long-range order,\cite{RLcrystal} momentum conservation is
replaced with crystal-momentum conservation. Thus, all that one needs
to do is to design a nonlinear photonic crystal (NPC)---whether
periodic or quasiperiodic---whose Fourier transform is peaked at all
the required mismatch wave vectors. Here we present the first
experimental realization of a device using this general solution: a
one-dimensional three-wave doubler.

Note that other schemes for multiple harmonic
generation have been demonstrated before.\cite{Ming_MultipleSHG2002,%
  Ming_FourColor2004,Fejer_MultipleHarmonics1999} Nevertheless, we
choose this relatively simple application of the LAB solution, as it
allows us to provide a detailed pedagogical description of the
approach. Other than demonstrating that the LAB solution indeed works,
we wish to clarify all the steps in the design process, so that others
will be able to implement it as well. We stress that the solution is
general and is not limited to such simple applications.

\section {Simultaneous Phase Matching of Several Interactions}

We consider second order nonlinear optical interactions in which three
beams couple through the nonlinear susceptibility $\chi^{(2)}$.  For a
planar process in which two constant undepleted beams, $E_1$ and
$E_2$, give rise to a third output beam $E_3$, its integrated field
amplitude is given by
\begin{equation}\label{eq:NonlinearInteraction}
E_3(\Dk)=\Gamma \int_A {g(\rv)\exp(i\Dk\cdot\rv)d^2r},
\end{equation}
where $A$ is the interaction area, and $\Gamma$ is a parameter
depending on the amplitudes of the incoming waves, on the indices of
refraction of all three waves, on the strength of the relevant
component of the nonlinear susceptibility tensor $d_{ij}$, and on the
interaction width $W$.  For example, for sum frequency generation in
MKS, $\Gamma= \omega_3^2 d_{ij} E_1 E_2/i c^2 k_3W$, where $c$ is the
speed of light in vacuum. The function $g(\rv)$ gives the spatial
dependence of the relevant nonlinear coupling coefficient, and ${\bf
  \Delta k}$ is the phase mismatch vector of the interacting waves.
For sum-frequency generation ${\bf \Delta k}$ would be ${\bf k}_1+{\bf
  k}_2-{\bf k}_3$.

It is clear from Eq.~(\ref{eq:NonlinearInteraction}) that the
intensity of the output beam is proportional to the Fourier spectrum
of the function $g(\rv)$, evaluated at the mismatch vector $\Dk$.
Thus, if we wish to simultaneously phase-match a set of $D$ three-wave
optical precesses, characterized by phase mismatch vectors ${\bf
  \Delta k}^{(j)}$, $j=1,\dots D$, we should design the spatial
structure of $g(\rv)$ so that its Fourier spectrum is peaked at all
the $D$ mismatch vectors. For a single process, the standard QPM
solution\cite{Armstrong1962,Fejer_QPM1992} is to design a
one-dimensional NPC with a period of $2 \pi /|\Delta k|$, for which
there is a first order Bragg peak in the spectrum at ${\bf \Delta k}$.
The LAB solution shows how to design an appropriate nonlinear photonic
crystal---whether periodic or quasiperiodic---such that its spectrum
contains Bragg peaks at any given set of $D$ mismatch vectors. The
approach that LAB adopt for this purpose is based on the so-called
dual-grid method, originally developed by de Bruijn\cite{deBruijn} and
later generalized\cite{GahlerRhyner_Equivalence,%
  RabsonHo_NonSymmorphic1988,RabsonHo1989} to become one of the
standard methods for creating tiling models of quasicrystals.

\begin{figure}[tb]
\includegraphics[width=\columnwidth]{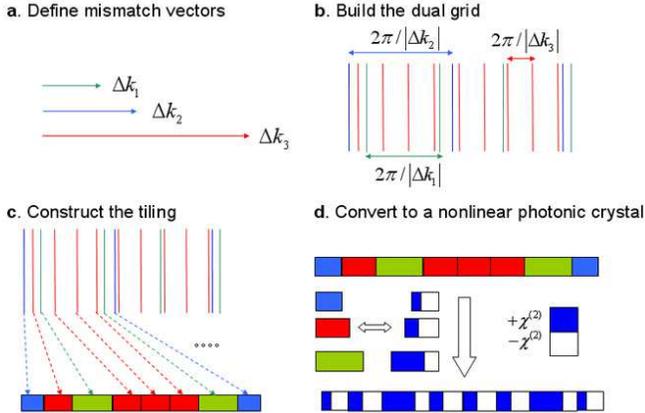}
\caption{{(Color online) Illustration of the LAB solution for
    designing a one-dimensional NPC for multiple collinear optical
    processes, using the dual grid method.} (a) The required mismatch
    vectors. (b) The dual grid, in which each family of lines is shown
    with a different color. (c) Tiling of the real-space line
    according to the order in which lines of different families appear
    in the dual grid. (d) Associating a given duty cycle with each
    tiling vector. Positively-poled segment are shown in blue, and
    negatively-poled segments are shown in white.}
\label{Fig:DGM_illustration}
\end{figure}

The reader in encouraged to consult LAB\cite{Lifshitz_PRL2005} for a
complete and rigorous treatment of the most general two-dimensional
multiple phase-matching problem, which we do not repeat here. Instead,
we give a detailed demonstration of the LAB solution in one dimension,
where all the optical processes are chosen to be collinear. In this case the
implementation of the dual-grid method for the design of the NPC, as
well as the experimental fabrication of the device, are relatively
simple. Nevertheless, we can still design nontrivial and interesting
devices, such as the three-wave doubler, implemented here. The basic
idea is to find a set of one-dimensional tiling vectors $a^{(i)}$,
$i=1,\dots D$, with which we can generate a one-dimensional tiling of
the line, whereby a tile is simply an interval on the line. We then
convert the tiling into an NPC by fabricating whole strips normal to
the tiles along the line.

Before starting we wish to point out that in special cases, the $D$
phase mismatch vectors $\Dk^{(j)}$ may be integrally dependent. This
means that one can use fewer than $D$ wave vectors to generate the NPC
and still have Bragg peaks at all $D$ points. It is then a matter of
choice whether to use the full set of dependent vectors---although, as
pointed out by LAB, it may be difficult in this case to control the
intensities of the peaks---or to prefer a smaller set of independent
vectors.  Here we keep all mismatch vectors and treat them as if they
were integrally independent.

\section{Designing a One-Dimensional Three-Wave Doubler}

We wish to design a device that will simultaneously phase-match three
collinear second-harmonic-generation processes, for three different
wavelengths in the fiber telecom C-band: $1530nm$, $1550nm$, and $1570nm$.
We choose to use the nonlinear crystal KTiOP$_4$ and operate at a
temperature of $100^\circ$C. At these conditions, the phase-mismatch
values for the three processes
are:\cite{KerenKTP1999,Emanueli_KTP2003} $\dk^{(1)}=.263 \mu m^{-1}$,
$\dk^{(2)}=.256 \mu m^{-1}$, and $\dk^{(3)}=.249 \mu m^{-1}$
respectively.  Thus, we need to design an NPC whose Fourier spectrum
contains three collinear wave vectors with these dimensions, as shown
schematically in Fig.~1a.

In what follows we describe the design of such a structure, as a
particular example of the LAB solution for $D$ collinear processes.
Generalizing from $D=3$ processes to an arbitrary number $D$ of
processes, follows directly by replacing all 3-component and
2-component vectors below by $D$-component and $(D-1)$-component
vectors, respectively. Generalizing to higher dimensional processes
requires use of the full solution, as described by LAB.

\subsection{Finding the tiling vectors}

To calculate the corresponding three collinear tiling vectors,
$a^{(i)}$, $i=1,\dots 3$, we first construct a single three-component
vector $\kv_1=(\Delta k^{(1)},\Delta k^{(2)},\Delta k^{(3)})$ from the
three given mismatch values.  This vector spans a one-dimensional
subspace of an abstract three-dimensional vector space. We complete it
to an orthogonal basis of the three-dimensional space by adding two
orthogonal vectors $\qv_2$ and $\qv_3$. These are, of course, not
unique, and we choose them to be $\qv_2=(0.6483,-0.3421,-0.3331)$ and
$\qv_3=(-0.3421,0.6672,-0.3240)$.  We use these three vectors as the
columns of a $3 \times 3$ non-singular matrix,
\begin{equation}\label{eq:Kmatrix}
\left( \begin{array}{ccc}
{\bf K}^{(1)} \\ {\bf K}^{(2)} \\ {\bf K}^{(3)}
\end{array} \right)
\equiv
\left( \begin{array}{ccc}
\Delta k^{(1)} & q_2^{(1)} & q_3^{(1)} \\
\Delta k^{(2)} & q_2^{(2)} & q_3^{(2)} \\
\Delta k^{(3)} & q_2^{(3)} & q_3^{(3)} \end{array} \right),
\end{equation}
whose rows $\Kv^{(j)}$, $j=1,\dots 3$, span the three-dimensional
vector space as well. We then find the three dual basis vectors,
denoted
\begin{equation}\label{eq:Amatrix2}
\left( \begin{array}{ccc}
{\bf A}^{(1)} \\ {\bf A}^{(2)} \\ {\bf A}^{(3)}
\end{array} \right)
\equiv
 \left( \begin{array}{ccc}
a^{(1)} & b_2^{(1)} & b_3^{(1)} \\
a^{(2)} & b_2^{(2)} & b_3^{(2)} \\
a^{(3)} & b_2^{(3)} & b_3^{(3)} \end{array} \right),
\end{equation}
by solving the three-dimensional orthogonality relations
\begin{equation} \label{eq:periodic_ortho}
{\bf A}^{(i)} \cdot {\bf K}^{(j)}=2\pi\delta_{ij}.
\end{equation}

Each row of the matrix (\ref{eq:Amatrix2}) is a dual-basis vector of
the form ${\bf A}^{(j)}=(a^{(j)},{\bf b}^{(j)})$. The $a^{(j)}$ are
the three required collinear tiling vectors, whose values are
calculated to be $a^{(1)}=8.3946\mu m$, $a^{(2)}=8.1652\mu m$, and
$a^{(3)}=7.95 \mu m$. The 2-dimensional vectors ${\bf b}^{(j)}$ can be
used, as explained by LAB, to analytically calculate the Fourier
transform of the NPC in order to determine the expected efficiencies
for the different nonlinear processes.

\begin{figure}[tb]
 \includegraphics[width=\columnwidth]{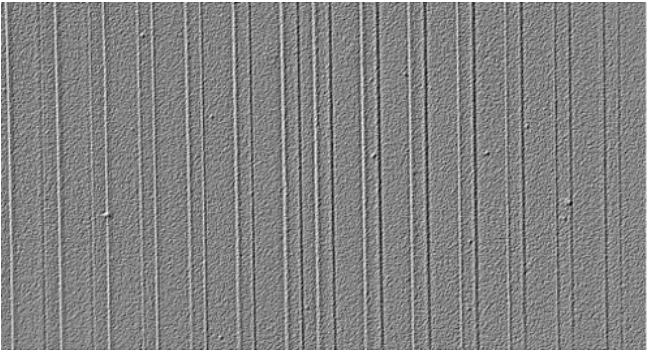}
\caption{{An optical microscope image of the demonstrated NPC}. The prominent
  elements correspond to the $a^{(1)}$ tiling vector, poled with a
  $100\%$ duty cycle. Their width is $8.5\mu m$. The distances between
  these elements are quasiperiodically ordered along with the
  $a^{(2)}$ and $a^{(3)}$ tiling vectors, whose widths are $8.1\mu m$ and
  $7.9\mu m$, respectively, and which are poled with a $0\%$ duty
  cycle.} 
 \label{Fig:Crystal_Pic}
\end{figure}

\subsection{Constructing the tiling}

If we were now asked to generate all points at integral linear
combinations of the three tiling vectors we would get the unwanted
outcome of a dense filling of the real line. To avoid this situation
we construct the dual grid, whose topology determines which of the
integral linear combination of the tiling vectors are to be included
in the one-dimensional tiling.

The dual grid is constructed by associating with each mismatch vector
$\Delta k^{(j)}$, $j=1,\dots3$, an infinite family of parallel lines
separated by a distance $L_j=2\pi / \Delta k^{(j)}$, as illustrated in
Fig.~1b. The set of all families together constitutes the dual grid.
We use the freedom to shift each family from the origin by an
arbitrary value of $f_j L_j$, where $0 \leq f_j<1$, so that lines from
different families never exactly coincide. Because the mismatch
vectors are independent over the integers, such a shift produces a
so-called gauge-transformation\cite{RabsonHo_NonSymmorphic1988,%
  RabsonHo1989} which has no effect on the resulting NPC.

The rest follows immediately, as the required order of the tiles in
the real-space structure is given by the order in which lines of
different families appear in the dual grid. This is illustrated in
Fig.~1c. This is the sense in which the topology of the dual grid
determines the real-space tiling in this trivial one-dimensional
setting. The duality is a statement that each line in the grid is
associated with a tile, or interval, in the tiling; and each interval
in the grid with a vertex of the tiling.  In our example,
approximately $800$ lines are required in each family to generate a
$1cm$ long one-dimensional nonlinear photonic quasicrystal.

\begin{figure}[pbt]
 \centerline{\includegraphics{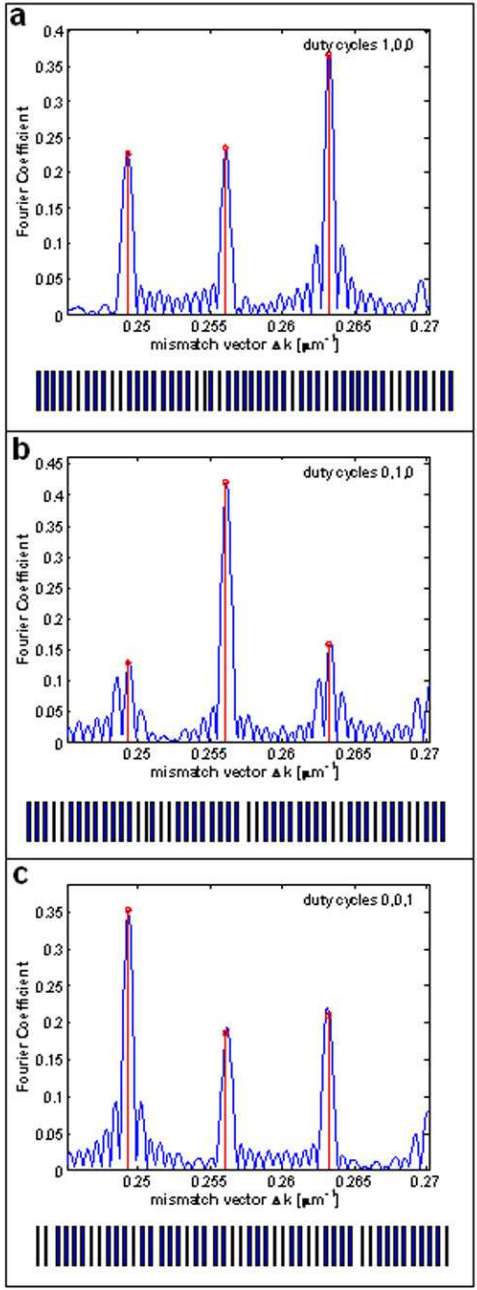}}
\caption{{(Color online) Spectral shaping}. Each panel shows the
  magnitude of the Fourier transform for a $1cm$ long NPC made to
  phase match the three collinear processes, described in the text. In
  each panel one of the tiling vectors is given a duty cycle of 100\%,
  denoted as 1, and the remaining two a duty cycle of 0\%, or 0. Each
  panel also shows a piece of the corresponding real-space
  representation of the NPC, where the smallest element size is $~8\mu
  m$.}
 \label{Fig:SpectrumShaping}
\end{figure}

\subsection{Building a nonlinear photonic crystal from the tiling}

To create an actual nonlinear photonic quasicrystal we modulate the
relevant component of the nonlinear susceptibility tensor $\chi^{(2)}$
according to the constructed tiling. Technology usually permits us to
use a binary modulation of $\chi^{(2)}$ so that the actual crystal can
be represented by a normalized function $g(r)=\pm1$. The simplest
representation from a theoretical standpoint would be to attach a thin
strip of value $g(r)=1$ to every vertex of the tiling, while assigning
the background a value of $g(r)=-1$. This is equivalent to a simple
convolution of the strip with delta functions at the vertices of the
tiling, and therefore gives the simplest analytical expression for the
Fourier transform of the function $g(r)$.\cite{Lifshitz_PRL2005}
Nevertheless, it does not necessarily produce an optimal NPC---one in
which the strongest Bragg peaks are associated with the three mismatch
vectors.  In general, one can use a numerical procedure in order to
optimize the required Bragg peaks.  See for example the treatment by
Norton and de Sterke.\cite{Norton_deSterke2003} Here we want to give a
few quick-and-simple solutions---in addition to the thin strips---that
are worth trying if one does not wish to deal with numerical
optimization.

A second option would be to use strips whose widths are equal to the
tile vectors and simply to change the sign of $g(r)$ from one strip to
the next. In this way, exactly half the tiles will give strips of value
$g(r)=1$, and the other half strips of value $g(r)=-1$. The generated
set of strips would be analogous to an antiferromagnetic
quasicrystal\cite{RLMagQC,RL1dAFM}, whose Fourier transform could also
be calculated analytically. We have found that this option does not
yield an optimal NPC for this application.

A third option, and the one which we actually implemented, is again to
use strips whose widths are equal to the tile vectors, but this time
associate a so-called duty cycle with each tiling vector.  This is
done by dividing each tiling vector into two segments, and assigning a
value of $g(r)=1$ to one segment, and $g(r)=-1$ to the other. The duty
cycle is the fraction of each strip with $g(r)=1$. This general
procedure is shown schematically in Fig.~1d. It gives us the ability
to perform simple spectral shaping. By varying the three duty cycles,
associated with the three tiling vectors, we can engineer the
magnitudes of the Fourier coefficients of the three required Bragg
peaks. What we actually find---contrary to common practice\cite{Keren_JQE1999,Keren_PRL2001,Ming_THG1997}---is that
the optimal NPC's are obtained when we use duty cycles of either 100\%
or 0\%. These are also easiest to fabricate in terms of the required
resolution as we associate a value of $g(r)=1$ or $g(r)=-1$ to tiles
as a whole.

In the experimental image of the NPC, shown in
Fig.~\ref{Fig:Crystal_Pic}, the tiling vector $a^{(1)}$ is associated
with strips of value $g(r)=1$, or a duty cycle of 100\%, while the
other two have a value of $g(r)=-1$, or a duty cycle of 0\%. In
Fig.~\ref{Fig:SpectrumShaping} we show numerical calculations of the
magnitude of the Fourier transform of $g(r)$, for the three possible
assignments of a value of $g(r)=1$ to one tiling vector, and a value
of $g(r)=-1$ to the other two. One clearly sees that in all cases
there are pronounced Bragg peaks exactly where we want them to be, but
the distribution of intensities changes as we vary the tiling vector
that is assigned a 100\% duty cycle.  The magnitudes of the Fourier
coefficients are comparable to the $2/\pi\simeq 0.6366$ figure of
merit, which is the magnitude of the first order Fourier component for
a one-dimensional periodic NPC. In fact, because the efficiency is
measured in terms of energy transfer it depends on the Fourier
intensity and on the square of the interaction length. Thus, the real
comparison should be with the efficiency per process of a sequence of
three periodic NPC's, each of length $L/3$, which is proportional to
$\left(\frac{1}{3}\cdot \frac{2}{\pi}\right)^2=0.045$.  For the
homogeneous nonlinear photonic quasicrystal that we implement here the
lowest process efficiency is proportional to $(1 \cdot 0.23)^2=0.053$
and the highest to $(1 \cdot 0.365)^2=0.133$, both better than the
composite periodic structure.

\begin{figure}[bt]
 \includegraphics[width=\columnwidth]{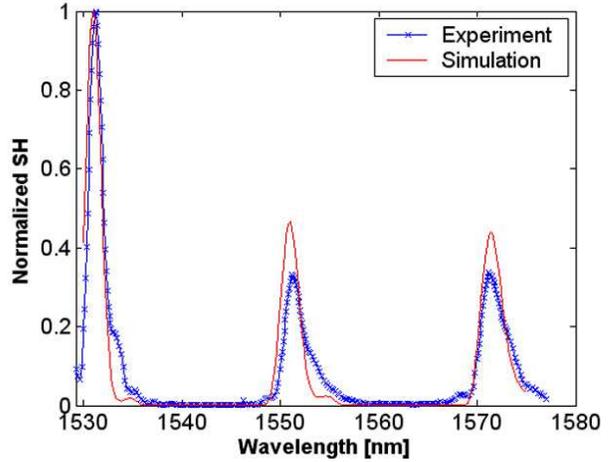}
\caption{{(Color online) Normalized pump wavelength response}. Experiment and
simulation results of second harmonic generation as a function of
pump wavelength. This figure corresponds to panel (a) of
Fig.~\ref{Fig:SpectrumShaping}.}  
 \label{Fig:ExpAndSim}
\end{figure}

\begin{table*}[tb]
\centering
\caption{Conversion efficiencies. Maximum conversion efficiencies are
  obtained for the indicated wavelengths. In the simulation, the
  maximum efficiency of the third process was shifted by $+0.4nm$.} 
\begin{tabular}{c c c c}\hline\hline
Wavelength & Experimental efficiency & Theoretical efficiency &
Simulated efficiency  \\ 
& [1/Watt] & [1/Watt] & [1/Watt] \\
\hline
1531$nm$ & 1.19e-3 & 1.5e-3 & 1.46e-3 \\
1551$nm$ & 3.37e-4 & 6.13e-4 & 6.83e-4  \\
1571$nm$ & 3.18e-4 & 5.49e-4 & 6.43e-4 \\
\hline
\end{tabular}
\label{table:efficiencies}
\end{table*}

The ability to shape the spectrum using different representation of
the function $g(r)$ can be very useful for cascaded processes that to
some extent are better performed in sequence. For example, in
multi-harmonic generation, where the outputs of some of the processes
serve as inputs for others, it might be beneficial to give precedence
to the efficiencies of the initial processes at the beginning of the
NPC and gradually, along the interaction length, change the balance in
favor of the latter processes. 

\section{Experimental Results}

The NPC was fabricated by electric field
poling\cite{Yamada_Poling1993} of KTiOPO$_4$, and is shown in
Fig.~\ref{Fig:Crystal_Pic}. The spatial modulation was performed along
$1cm$ with a $2\times 1 mm^2$ cross section.  To test the NPC we used
a pump beam from a tunable external cavity diode laser, followed by an
Erbium-doped fiber amplifier, and a fiber polarization controller
whose purpose is to polarize the beam to be perpendicular to the plane
of modulation of the NPC. The pump beam was then chopped at a
frequency of 1kHz and focused to a waist of $20\mu m$ in the middle
of the crystal. Its power was varied during the experiment in the
range of $4-40mW$.  The NPC was kept at a constant temperature of
$100^\circ$C. The input pump wavelength was varied in the range
$1528nm-1577nm$ and the resulting second harmonic power was measured
using a calibrated Silicon photo-diode, followed by a lock-in
amplifier.  The results were compared with a simulation employing a
split step Fourier method,\cite{Agrawal} in which a non-depleted
Gaussian beam was used as a pump. Both results exhibit excellent
agreement, as can be seen in Fig.~\ref{Fig:ExpAndSim}, and show that
this device is indeed most efficient for second harmonic generation of
the three designated pump wavelengths.

Note that for focused Gaussian
beams\cite{BoydKleinman_FocusedBeams1968} the actual peak efficiencies
are shifted by $1nm$ towards longer wavelengths, as compared to the
idealized design using plane wave interaction.  However, with
temperature tuning the results can be shifted in both directions. For
the wavelength range used here, the shift rate is $0.1nm$ per degree
centigrade. The conversion efficiencies were also calculated
theoretically using the Boyd-Kleinman formalism for focused Gaussian
beams, propagating within a homogeneous nonlinear
crystal,\cite{BoydKleinman_FocusedBeams1968} modified to include the
relevant Fourier coefficient. The nonlinear conversion efficiencies
for the wavelengths for which conversion was maximal are shown in
Table~\ref{table:efficiencies}. The small discrepancies between theory
and simulation are attributed to the assumption, made in the theoretical
calculation, that each processes is affected only by its relevant
Fourier component. In the simulation the propagating waves experience
small but still non-negligible contributions from close-by spectral
components.

\section{Summary}

We present the first experimental realization of a nonlinear optical
device employing the general solution of the phase matching problem,
given by LAB. The demonstrated device is a one-dimensional three-wave
doubler. We show that by simply using $100\%$ or $0\%$ duty cycles,
each of the three wave doubling processes exhibits high efficiency.
Moreover, the efficiencies are all higher than for a device employing
a periodic modulation with the same overall interaction length. We
demonstrate the ability to perform spectral shaping of the response by
changing the duty cycles, associated with the different tiles. This
allows us to strengthen certain processes at the expense of others.
It should be stressed once more that the LAB solution is general and
can support any set of nonlinear processes, not only for
one-dimensional problems but also in two or three dimensions, without
any symmetry restrictions. In addition, this general method can be
applied to a broad range of problems from generation of radiation
sources through all-optical processing to generation of entangled
photons in quantum optics applications.

\section{Acknowledgments}

This research is supported by the Israel Science Foundation through
grants no.~960/05 and 684/06.

\end{document}